\documentclass[10pt,twocolumn]{revtex4-1}

\usepackage{hyperref}
\usepackage{amsmath}  
\usepackage{amsfonts} 
\usepackage{graphicx} 
\usepackage[utf8]{inputenc}
\usepackage{xcolor}
\usepackage{gensymb}
\usepackage[english]{babel}
\begin{document}

\title{Cause-effect relationships in Maxwell's equations and their implications in the teaching of electromagnetism in introductory physics courses}

\author{Álvaro Suárez}
\email{alsua@outlook.com} 
\affiliation{Departamento de Física, Consejo de Formación en Educación, Montevideo, Uruguay}

\author{Arturo C. Martí}
\email{marti@fisica.edu.uy}
\affiliation{Instituto de F\'{i}sica, Facultad de Ciencias, Universidad de la Rep\'{u}blica,
             Igu\'{a} 4225, Montevideo, 11200, Uruguay}

\author{Kristina Zuza}
\email{kristina.zuza@ehu.eus} 
\affiliation{Department of Applied Physics, University of Basque Country, Spain}         

\author{Jenaro Guisasola}
\email{jenaro.guisasola@ehu.eus} 
\affiliation{Department of Applied Physics, University of Basque Country, Spain}    

\date{\today}

\begin{abstract}
\end{abstract}

\begin{abstract}
A thoughtless treatment of Maxwell's equations can lead to the interpretation of the existence of a causal relationship between their different terms  and, therefore, that an electric field that varies in time generates a magnetic one and vice versa. In this article we address the problems associated with these interpretations and their consequences for the teaching of physics in introductory university physics courses. First, we develop the main arguments that support that charge and current densities, constant or variable in time, are the generators of electric and magnetic fields. Then, we propose a number of classroom examples at the level of introductory courses that allow us to discuss the reasons why considering the electric and magnetic fields as disjoint entities leads to contradictions. Finally, we analyze their implications in introductory physics teaching.\\

\textbf{About this paper} \\
This paper is an English translation of the article “\textit{Las relaciones causa-efecto en las ecuaciones de Maxwell y sus implicancias en la enseñanza del electromagnetismo en los cursos introductorios de Física}", originally published in the Revista Brasileira de Ensino de Física (Suárez, A., Martí, A. C., Zuza, K., \& Guisasola, J., 2022, 44, e20220230). Readers are encouraged to cite the original publication when referencing this work. The original version is available \href{https://www.scielo.br/j/rbef/a/9wzLbC7smz6qvnwvqqgzTCx/?lang=es}{here}. 

\end{abstract}

\maketitle 

\section{Introduction}
The unified description of the electromagnetism presented by J. C. Maxwell \cite{maxwell1873treatise} in his \textit{“A Treatise on Electricity and Magnetism"} is one of the most important scientific constructs in physics. There is an undeniable need for students at different educational levels to acquire an adequate understanding of the concepts involved. Considerable research in the field of Physics Education Research (PER) has addressed the teaching of electromagnetism in regard to conceptual understanding, problem solving, and curriculum and instruction \cite{docktor2014synthesis}. However, the advances in PER have generally had less impact in introductory physics textbooks than in other areas such as mechanics \cite{galili2005energy,chabay2006restructuring,zuza2012rethinking}.

One aspect in particular that has been discussed extensively in the literature over the last 50 years is the sources of electric and magnetic fields. A thoughtless treatment of Maxwell's equations in introductory physics textbooks usually leads to the interpretation that electric fields can be generated by charged particles or time-varying magnetic fields, and, at the same time, that magnetic fields are produced by currents or time-varying electric fields \cite{serway2018physics,halliday2003physics}. However, studies on the nature of classical electrodynamics show that constant and time-varying charge and current densities are the generators of electromagnetic fields \cite[pp~516-517]{jefimenko1989electricity}\cite{jefimenko2004presenting}\cite[pp~48-49]{rosser2013interpretation}\cite[pp~261-262]{heald2012classical}\cite[pp~449-450]{griffiths2013introduction}. This approach has had little impact on introductory physics textbooks, neglecting these aspects could lead to students inadequately interpreting the sources of electromagnetic fields and the meaning of Maxwell's equations \cite{rainson1994students,guisasola2004difficulties,hill2010rephrasing,hill2011reanalyzing,campos2019}. In this article we address, at an introductory physics level, the problem of the sources of the fields and cause-effect relationships in Maxwell's equations and present a series of situations in which the important role of a unified treatment of the electromagnetic field is shown. 

In the following section, we develop the main arguments that support that charge and current densities, constant or variable in time, are the sources of electric and magnetic fields. In the next section, we show the phenomenon of the displacement current and the relevance of the sources of the electromagnetic field for its interpretation in the framework of classical electromagnetic theory. Then, we present a series of examples at the undergraduate introductory physics level that show how misinterpretations can be avoided with a unified treatment of the fields and the emphasis on their sources, which contributes to a better understanding of Maxwell's equations. Finally, we conclude with a discussion and implications for teaching.

\section{\label{Sources}Sources of electromagnetic fields}

In their integral form, Maxwell's equations, the formalism that allows us to predict the time evolution of electromagnetic fields, are expressed as follows: 
\begin{eqnarray}
\label{gaussE}
\oint \mathbf{E} \cdot \mathbf{d A}=\frac{q}{\varepsilon_{0}}
\end{eqnarray}
\begin{eqnarray}
\label{gaussB}
\oint \mathbf{B} \cdot \mathbf{d A}=0
\end{eqnarray}
\begin{eqnarray}
\label{faraday}
\oint \mathbf{E} \cdot \mathbf{d l}=-\frac{d}{d t} \int \mathbf{B} \cdot \mathbf{d A}
\end{eqnarray}
\begin{eqnarray}
\label{maxwell}
\oint \mathbf{B} \cdot \mathbf{d l}=\mu_{0} I+\mu_{0} \varepsilon_{0} \frac{d}{d t} \int \mathbf{E} \cdot \mathbf{d A}
\end{eqnarray}
where \textbf{E} is the electric field, \textbf{B} the magnetic field, $q$ the electric charge, $I$ the conduction current intensity, and $\varepsilon_{0}$ and $\mu_{0}$ the electric permittivity and the magnetic permeability of free space respectively. Maxwell's equations must be interpreted as a rule for calculating the fields from the charge distributions. This interaction also occurs in the opposite direction as the fields exert influence on the charges and currents through the Lorentz force.

Generally, in introductory physics courses, when electric and magnetic fields are discussed, it is common to start with Gauss's law, Eq. 1, and the stationary implications of Ampere-Maxwell's and Faraday's laws, Eqs.~\ref{faraday}-\ref{maxwell}. These laws allow us to describe static fields and they reveal that charged particles generate electric fields, whereas electric currents generate magnetic fields. This framework presents the first obstacles for students to adequately interpret the sources of the fields \cite{campos2021}. 
Regarding Gauss's law for the electric field, all the charges in the space contribute to the electric field and must be considered for the calculation of the flux through the closed surface and  not only those enclosed by the  surface \cite{wangsness1986electromagnetic}.  
On the other hand, according to Ampère's law, the magnetic field used to determine the circulation along a closed curve is the one corresponding to all the conduction currents and not only to those enclosed by the curve. Guisasola, et al \cite{guisasola2003analisis,guisasola2008gauss} found that, when asking students which charges generate the electric field of an infinite plane obtained by Gauss's law, many understand that only the charges enclosed by the Gaussian surface contribute. Similarly, when students are shown a typical Amperian loop to find the magnetic field of an infinite solenoid and asked about the currents that generate the field, many believe that only those enclosed by the curve contribute. These results show that students frequently exhibit a lack of understanding about the nature of the relationships between the different elements of Gauss's and Ampère's laws,  they do not adequately consider all variables and  they simplify cause-effect relationships \cite{guisasola2003analisis,guisasola2008gauss}. 

When studying situations with time-dependent fields, Faraday's and Ampère-Maxwell's laws come into play, as they describe non-electrostatic electric fields and magnetic fields associated with displacement currents. In these equations, the fields are related to each other through the terms of the time derivatives of the flows \cite{alonso1983fundamental}. In this scenario, we find another obstacle to properly interpreting the relationships between the different variables in Maxwell's equations and the sources of the fields, which, as we will see, could originate in the way in which the terms that relate them are interpreted.

An incomplete analysis of Faraday's and Ampère-Maxwell's laws would allow us to infer that a time-varying magnetic field generates an electric field and vice versa. This is the dominant interpretation in many introductory physics textbooks. For example, in the case of Faraday's law, Tipler and Mosca \cite[p~1031]{tipler2007physics} state that “According to Faraday’s law, a changing magnetic flux produces an electric field whose line integral around a closed curve is proportional to the rate of change of magnetic flux through any surface bounded by the curve". Serway and Jewett \cite[p~809]{serway2018physics} conclude that “Equation 30.8 ($\oint \mathbf{E} \cdot \mathbf{d} \mathbf{l}=-d \Phi_{\mathrm{B}} /dt$) is the general form of Faraday’s law. It represents all situations in which a changing magnetic field generates an electric field". Resnick, Halliday and Krane \cite[p~784]{halliday2003physics}, on the other hand, argue that “It is in this form that Faraday's law appears as one of the four basic Maxwell equations of electromagnetism. In this form, it is apparent that Faraday's law implies that a changing magnetic field produces an electric field". Similar interpretations can be found in the literature regarding Ampère-Maxwell's law. Tipler and Mosca \cite[p~1031]{tipler2007physics} assert that “We thus have the interesting reciprocal result that a changing magnetic field produces an electric field (Faraday’s law) and a changing electric field produces a magnetic field (generalized form of Ampère’s law)". Serway and Jewett \cite[p~876]{serway2018physics} argue that Ampère-Maxwel's law “describes the creation of a magnetic field by a changing electric field and by electric current".  Finally, Resnick, Halliday and Krane \cite[p~862]{halliday2003physics} argue that “a magnetic field is set up by a changing electric field". 

We can also find the idea that a time-varying electric field generates a magnetic field and vice versa in other introductory physics textbooks  \cite[p~940]{bauer2013university}\cite[p~945]{fishbane2005thornton} \cite[pp~773 and 813]{giancoli2008physics} 
\cite[pp~852 and 885]{knight2017physics} 
\cite[pp~994 and 1073]{ohanian2007physics} 
\cite[pp~968 and 972]{young2019university}.
According to Bunge \cite[p.~62]{bunge2009causality}, the mentioned interpretations could be based on the restricted version of the principle of causality in agreement with the principle of \textit{delayed action}. This interpretation of the causality as
delayed action explain that “there is always a time delay between the cause and the effect, the former being prior in time to the latter, so that (relatively to a given physical system, such as a reference system), C and E cannot be both distant in space and simultaneous".

The analysis on the generation of electric and magnetic fields from their sources carried out by Jefimenko \cite{jefimenko1989electricity}, clarified the question about the cause-effect relationships and the timing of the electric and magnetic field. This analysis concludes that the general expressions for the fields are given by
 \cite[pp~516-517]{jefimenko1989electricity}
\begin{eqnarray}
\label{jefimenkoE}
\mathbf{E}=\frac{1}{4 \pi \varepsilon_{0}}\int\left(\frac{\rho\left(r^{\prime}, t^{\prime}\right)}{\left|\mathbf{r}-\mathbf{r^{\prime}}\right|^{3}}+\frac{\dot{\rho}\left(r^{\prime}, t^{\prime}\right)}{c\left|\mathbf{r}-\mathbf{r^{\prime}}\right|^{2}}\right)\left(\mathbf{r}-\mathbf{r^{\prime}}\right) d v^{\prime}\nonumber
\end{eqnarray}
\begin{eqnarray}
-\frac{1}{4 \pi \varepsilon_{0}}\int \frac{\dot{\mathbf{J}}\left(r^{\prime}, t^{\prime}\right)}{c^{2} \mid \mathbf{r}-\mathbf{r^{\prime} \mid}} d v^{\prime}
\end{eqnarray}
\begin{eqnarray}
\label{jefimenkoB}
\mathbf{B}=\frac{\mu_{0}}{4 \pi} \int\left(\frac{\mathbf{J}\left(r^{\prime}, t^{\prime}\right)}{\left|\mathbf{r}-\mathbf{r^{\prime}}\right|^{3}}+\frac{\dot{\mathbf{J}}\left(r^{\prime}, t^{\prime}\right)}{c\left|\mathbf{r}-\mathbf{r^{\prime}}\right|^{2}}\right) \times\left(\mathbf{r}-\mathbf{r^{\prime}}\right) d v^{\prime}
\end{eqnarray}
where \textbf{E} and \textbf{B} are evaluated at position $\mathbf{r}$ at time instant \textit{t}, being ${\textbf{r}}^{\prime}$ the distance from the origin of coordinates to charge density $\rho$ and current density \textbf{J}, $c$ is the speed of light and $t^{\prime}=t-\left|{\textbf{r}}-{\textbf{r}^{\prime}}\right| / c$. As expected, the Jefimenko equations are reduced to Coulomb's and Biot-Savart laws for a moving point charge when the charge distributions are at rest and the currents are constant.

According to this analysis, the sources of the fields are the constant or time-varying charge and current densities. To clarify this matter, let us consider an electromagnetic wave propagating in space, with the electric and magnetic fields in phase. The cause of the wave is a distribution of charges oscillating, with electric and magnetic fields at a certain position \textbf{r} and time $t$ linked to the movement of these charges over a previous time $t^{\prime}$. It impossible to affirm the existence of an \textit{ electric wave} that subsequently produces a magnetic field.

In contemporary classical electromagnetic theory, the electric and magnetic fields conform a single object, the electromagnetic field \cite{feynman1963feynman}. Thus, to conceive that if we are given one field, we can obtain the evolution of the other ignores the fact that they are components of the same entity and, therefore, they cannot interact with each other \cite{jefimenko2004presenting}. In this sense, Faraday's and Ampère-Maxwell's laws cannot imply cause-effect relationships, since they link two quantities that are simultaneous and, therefore, neither of these quantities can be the source of the other \cite{hill2010rephrasing,hill2011reanalyzing}. 

\section{\label{displacement}The case of the displacement current}

Let us now consider the case of the displacement current and how to interpret its possible role as a source of magnetic fields. To do this, let us analyze the current density term in the Jefimenko equation for the magnetic field. In Eq.~\ref{jefimenkoB}, \textbf{J} includes, in addition to the current density of free charges, the polarization current density $\partial \mathbf{P} / \partial t$ and the magnetization current density $\nabla \times\mathbf{M}$  \cite{griffiths1991time,jefimenko1992solutions}, being \textbf{P} and \textbf{M} the polarization and magnetization vectors, respectively. Although it may come as a surprise, the density of the vacuum displacement current $\left(\varepsilon_{0} \partial \mathbf{E} / \partial t\right)$ is not a source of magnetic field \cite{Rosser1976}. To properly understand this result, we must go back to Maxwell's early work.

The displacement current was first introduced by James Clerk Maxwell in \textit{“On Physical Lines of Force"}, published in 1861, where he succeeded in developing an electromagnetic theory based on a mechanical model of the ether \cite{chalmers1975maxwell, lima2019surgimento}. Thanks to the introduction of this contribution he obtained a version of the continuity equation similar to the one currently used \cite[p~496]{maxwell1890scientific}. However, it is worth mentioning that he considered it similar in nature to conduction current
 \cite[p~161]{darrigol2003electrodynamics}. In that same work, Maxwell deduced for the first time the speed of propagation of electromagnetic waves through hypotheses linked to the mechanism of the electromagnetic field, suggesting that “light consists in the transverse undulations of the same medium which is the cause of electric and magnetic phenomena" \cite[p~500]{maxwell1890scientific}. 

Aware of the limitations and difficulties associated with his mechanical model of the ether, he decided to make it independent of the electromagnetic field. Thus, in 1864, he published \textit{“A Dynamical Theory of the Electromagnetic Field"}, in which he presents eight equations of the electromagnetic field and an electromagnetic theory of light that can be tested experimentally \cite{berkson2014fields}. It is in this work that he clearly describes his view of the physical meanings of electrical displacement and displacement current. “Electrical displacement consists in the opposite electrification of the sides of a molecule or particle of a body which may or may not be accompanied with transmission through the body… The variations of the electrical displacement must be added to the currents p, q, r to get the total motion of electricity…” \cite[p~554]{maxwell1890scientific}. Two fundamental conclusions can be drawn from this fragment: on the one hand, that in \textit{A Dynamical Theory of the Electromagnetic Field} Maxwell considers the displacement current as another type of current that contributes to the total current \cite{chalmers1975maxwell,darrigol2003electrodynamics} and, on the other hand, that the electric displacement was for him what the polarization vector is for us today \cite{arthur2009elementary}.  

In 1873, Maxwell published his main work, \textit{“A Treatise on Electricity and Magnetism"}, in which he presents in detail the whole electromagnetic theory. There he clearly states his position regarding the temporal variation of the electric displacement as current, asserting that this magnitude must generate a magnetic
field, just like conduction currents: “The current produces magnetic phenomena in its neighborhood… We have reason for believing that even when there is no proper conduction, but merely a variation of electric displacement, as in the glass of a Leyden jar during charge or discharge, the magnetic effect of the electric movement is precisely the same"\cite[pp~144-145]{maxwell1873treatise}. From the above, we see that, from Maxwell's point of view, the displacement current generates a magnetic field, just like the conduction current. Moreover, given Maxwell's conception of space, in particular his conviction about the existence of the ether, he considers the displacement current a consequence of the variation of the electric displacement in any mechanical medium, and therefore it is always associated with a movement of bound charges, unlike the currently accepted view. 

The suppression of the mechanical ether presents several problems for the interpretation of the displacement current, especially when analyzing the problems of the generation and propagation of electromagnetic fields in a vacuum. The absence of the ether makes the vacuum displacement current a simple term directly proportional to the rate of change of the electric field \cite{Roche_1998}. To understand this, it should be noted that when a medium is polarized by the effect of an electric field, the electric displacement \textbf{D} is given by: 
\begin{eqnarray}
\label{desplazamiento}
\mathbf{D}=\varepsilon_{0} \mathbf{E}+\mathbf{P}
\end{eqnarray}
whereas the displacement current results as follows:
\begin{eqnarray}
\label{corrientedesplazamiento}
\frac{\partial \mathbf{D}}{\partial t}=\varepsilon_{0} \frac{\partial \mathbf{E}}{\partial t}+\frac{\partial \mathbf{P}}{\partial t}
\end{eqnarray}
It should also be noted that while the term $\partial \mathbf{P} / \partial t$ is associated with a real motion of bound charges, the last term, $\varepsilon_{0} \partial \mathbf{E} / \partial t$, corresponds to the vacuum's contribution to the displacement current. Therefore, and contrary to the view of Maxwell, we can have a displacement current in a vacuum. This last point is crucial to understand why the vacuum displacement current is not a source of magnetic fields. 
A variation of the vector \textbf{D} in vacuum could be attributed solely to a variation of the electric field, and, as we discussed in the previous section, a time-varying electric field is not a source of magnetic fields \cite{Rosser1976,french2000maxwell}. 

\section{\label{Maxwell}Maxwell's equations imply relationships between the fields}

Through the analysis of some electromagnetic phenomena at the level of introductory physics in this section we highlight that Maxwell's equations establish relationships between different magnitudes at the same instant of time. This analysis will allow us to recognize that considering the electric and magnetic fields as disjoint entities leads to contradictions.

\subsection{\label{chargedparticle}A moving point charge}

\begin{figure}[h!]
\centering
\includegraphics{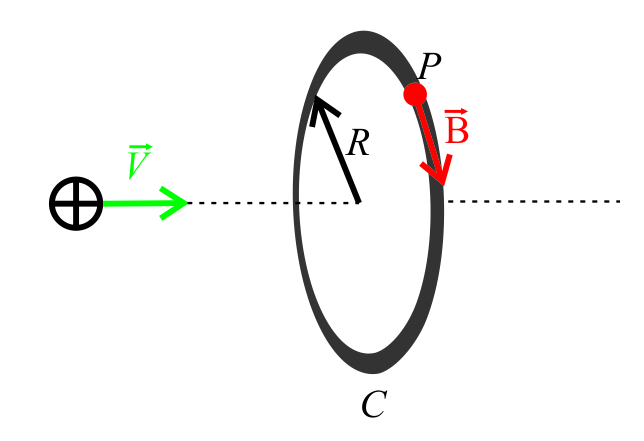}
\caption{Point charge performing a URM with respect to an inertial frame of reference $R$ and a closed curve $C$.}
\label{figure1}
\end{figure}

An electromagnetic phenomenon discussed in almost all introductory physics textbooks for science and engineering is the magnetic field generated by a point charge in uniform rectilinear motion. When the point charge moves with constant velocity, it generates electromagnetic fields around it. The magnetic field generated at a point P in space at a distance $r$ from the particle is usually calculated by means of the Biot-Savart law for a moving point charge.
\begin{eqnarray}
\label{biot}
\mathbf{B}=\frac{\mu_{0} q \mathbf{v} \times \mathbf{r}}{4 \pi r^{3}}.
\end{eqnarray}
However, it is also possible to obtain the magnetic field using the Ampère-Maxwell law (Eq.\ref{maxwell})
\cite{buschauer2013derivation,suarez2013campo}. Assume a closed curve $C$ of radius $R$ that passes through point P as shown in figure~\ref{figure1}. 
If we apply equation \ref{maxwell}, the term $\mu_{0}I$ vanishes since no conduction current passes through the surface $S$ delimited by the curve $C$. Therefore, there is a displacement current that appears because the point charge is approaching curve C and the flux of electric field is getting larger. In this way, the magnetic field circulation along the curve $C$, is given by
\begin{eqnarray}
\label{10}
\oint \mathbf{B} \cdot \mathbf{dl}=\mu_{0} \varepsilon_{0} \frac{d}{d t} \int \mathbf{E} \cdot \mathbf{dA}=\mu_{0} \varepsilon_{0} \frac{d \Phi_{E}}{d t}
\end{eqnarray}
where $\varepsilon_{0} \frac{d \Phi_{E}}{d t}= I_D$ is the displacement current through the surface bounded by curve $C$. 

\begin{figure}[h!]
\centering
\includegraphics{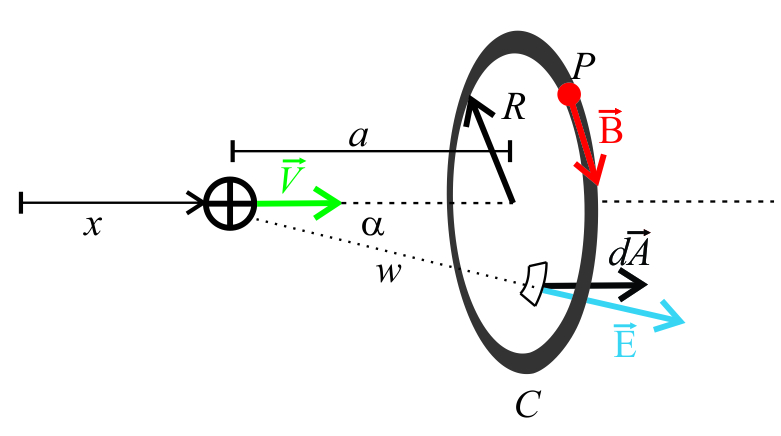}
\caption{Geometric configuration for the calculation of the displacement current of example A.}
\label{figure2}
\end{figure}

If the velocity of the point charge is much less than the speed of light, the electric flux through a surface element \textbf{dA}, is given by
\begin{eqnarray}
\label{11}
d \Phi_{E}=\mathbf{E} \cdot \mathbf{dA}=\frac{q}{4 \pi \varepsilon_{0} w^{2}} \cos \alpha d A
\end{eqnarray}
where \textit{w} is the distance from the point charge to the surface element and $\alpha$ is the angle formed by these vectors, as shown in figure~\ref{figure2}.
Since $\cos\alpha=a/w$ and $w=\sqrt{a^{2}+h^{2}}$, where \textit{h} is the distance from the center of the curve to the element of integration, if we substitute in Eq.~\ref{11} and integrating over the entire surface results
\begin{eqnarray}
\label{12}
\Phi_{E}=\frac{q}{4 \pi \varepsilon_{0}} \int_{0}^{R} \int_{0}^{2 \pi} \frac{a}{\left(a^{2}+h^{2}\right)^{3 / 2}} h d \varphi d h,
\end{eqnarray}
which can be integrated to obtain 
\begin{eqnarray}
\label{13}
\Phi_{E}(a)=\frac{q}{2 \varepsilon_{0}}\left[1-\frac{a}{\left(a^{2}+R^{2}\right)^{1 / 2}}\right].
\end{eqnarray}

From the temporal derivative of the flux the displacement current can be obtained as
\begin{eqnarray}
\label{14}
I_{D}=\varepsilon_{0} \frac{d \Phi_{E}}{d t}=\varepsilon_{0} \frac{d \Phi_{E}}{d a} \frac{d a}{d t}=\varepsilon_{0} \frac{d \Phi_{E}}{d a}(-v),
\end{eqnarray}
where $v=\frac{d x}{d t}=-\frac{d a}{d t}$.
If we substitute Eq.~\ref{13} in Eq.~\ref{14} and derive results
\begin{eqnarray}
\label{15}
I_{D}=\frac{v q}{2} \frac{R^{2}}{\left(a^{2}+R^{2}\right)^{3 / 2}}
\end{eqnarray}
which expresses the displacement current through the surface $S$ bounded by the curve $C$. To obtain the magnetic field, we substitute Eq.~\ref{15} in Eq.~\ref{10}
\begin{eqnarray}
\label{16}
\oint \mathbf{B} \cdot \mathbf{dl}=\frac{\mu_{0} v q}{2} \frac{R^{2}}{\left(a^{2}+R^{2}\right)^{3 / 2}}.
\end{eqnarray}
Using that the magnitude of the magnetic field is constant over the curve $C$, it is easily obtained
\begin{eqnarray}
\label{19}
B=\frac{\mu_{0} v q}{4 \pi} \frac{R}{r^{3}}=\frac{\mu_{0} v q}{4 \pi} \frac{1}{r^{2}} \frac{R}{r}
\end{eqnarray}
Writing Eq.~\ref{19} depending on the angle formed between \textbf{\textit{r}} and \textit{\textbf{v}} we obtain
\begin{eqnarray}
\label{20}
B=\frac{\mu_{0} v q}{4 \pi} \frac{\operatorname{sen} \theta}{r^{2}}
\end{eqnarray}
which can be expressed vectorially in the following form
\begin{eqnarray}
\label{21}
\mathbf{B}=\frac{\mu_{0} q \mathbf{v} \times \mathbf{r}}{4 \pi r^{3}}.
\end{eqnarray}
Being the last equation identical to the Biot-Savart law for a moving point charge. At this point, we must be careful with the physical interpretation of the phenomenon. We are faced with a situation where we can determine the magnetic field at a point in space directly by the Biot-Savart law for a moving point charge or by Ampère-Maxwell's law. If we assign a cause-effect relationship to both laws and analyze the situation applying the former, we would say that, because the point charge is in motion, it generates a magnetic field around it, whereas if we apply the latter, we would conclude that the time-varying electric field is the cause of the magnetic field. If both interpretations were valid we would have to consider both contributions for the calculation of the magnetic field \cite{Roche_1998}.

When a point charge is moving with constant velocity it generates an accompanying electromagnetic field. It is not accurate to say that the electric field generates the magnetic field, because both are parts of the same object. As we mentioned earlier, the sources of electric and magnetic fields are the charge and current densities, either constant or time-varying.

\subsection{\label{capacitor}Ampère-Maxwell's law and the charge of a capacitor}

\begin{figure}[h!]
\centering
\includegraphics{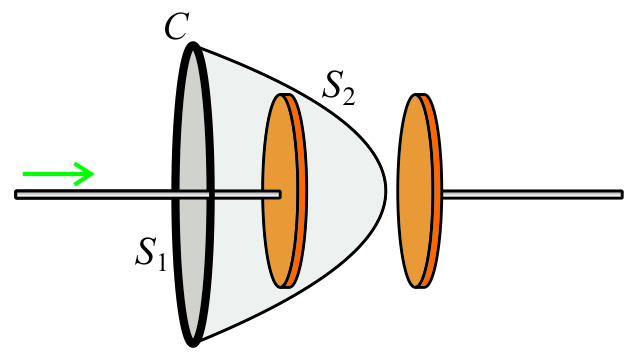}
\caption{Surfaces $S_{1}$ and $S_{2}$ are delimited by the same curve $C$.}
\label{figure3}
\end{figure}

Consider the best-known example for introducing the displacement current, showing a capacitor being charged, a closed curve $C$, and two surfaces $S_{1}$ and $S_{2}$ as shown in figure \ref{figure3}.
A typical interpretation of this problem is that depending on the surface taken for the calculation of the magnetic field circulation, the current generated by the magnetic field changes. For example, Resnick, Halliday and Krane \cite[p~862]{halliday2003physics} argue that “In the first case, it is the current through the surface $\left(S_{1}\right)$ that gives the magnetic field, and in the second case, it is the changing electric flux through the surface $\left(S_{2}\right)$ that gives the magnetic field”. 

The actual source of a magnetic field at a point cannot depend on the surface taken to calculate circulation. In that sense, assertions such as the one transcribed above may lead students to believe that the magnetic field appearing in the line integral of Ampère-Maxwell's law is due only to the currents crossing the surface, which may result in cause-effect interpretations of Maxwell's equations \cite{hill2011reanalyzing}. 

\begin{figure}[h!]
\centering
\includegraphics{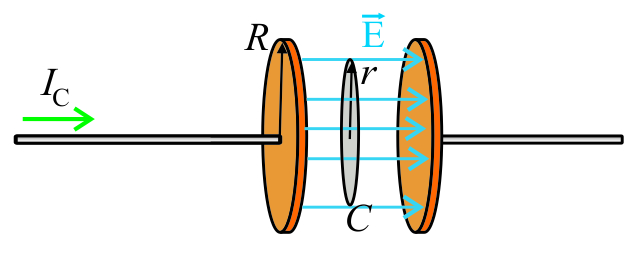}
\caption{Representation of the electric field between the plates of the capacitor and a closed curve $C$.}
\label{figure4}
\end{figure}

As in the previous example, other analyses could also lead to conceptual errors among students, promoting causal linear reasoning in the interpretation of Ampère-Maxwell's law. Consider for example the determination of the magnetic field at a point between the plates of a circular capacitor being charged. To find the magnetic field at a distance $r \leq R$ from the axis of symmetry, where $R$ is the radius of the plates, we can apply the Ampère-Maxwell law to the closed curve $C$ indicated in figure \ref{figure4}. 

While a conduction current $I_{C}$ flows through the cable, there is only displacement current on the surface $S$ delimited by the curve $C$. Therefore, the magnetic field circulation along the curve $C$ turns out to be
\begin{eqnarray}
\label{22}
\oint \mathbf{B} \cdot \mathbf{dl}=\mu_{0} I_{D n e t a}
\end{eqnarray}
where $I_{D neta}$ is displacement current through the surface $S$ given by
\begin{eqnarray}
\label{23}
I_{D n e t a}=\varepsilon_{0} \frac{d}{d t} \int_{0}^{r} \mathbf{E} \cdot \mathbf{dA}.
\end{eqnarray}
If we assume that for any instant of time the electric field between the plates is uniform and outside it is zero, the net displacement current through the surface $S$ can be expressed as a function of the total displacement current $I_{D}$ and the radius $R$ of the capacitor plates
\begin{eqnarray}
\label{24}
I_{D n e t a}=\frac{r^{2}}{R^{2}} I_{D}.
\end{eqnarray}
Substituting Eq.~\ref{24} in Eq.~\ref{22}
\begin{eqnarray}
\label{25}
\oint \mathbf{B} \cdot \mathbf{dl}=\mu_{0} \frac{r^{2}}{R^{2}} I_{D},
\end{eqnarray}
as the magnitude of the magnetic field is constant on the curve $C$
\begin{eqnarray}
\label{26}
B 2 \pi r=\mu_{0} \frac{r^{2}}{R^{2}} I_{D}
\end{eqnarray}
we obtain the magnetic field at a distance $r \leq R$ from the center of the capacitor
\begin{eqnarray}
\label{27}
B=\frac{\mu_{0} r}{2 \pi R^{2}} I_{D}.
\end{eqnarray}
On the other hand, since the total displacement current is equal to the conduction current, it results
\begin{eqnarray}
\label{28}
B=\frac{\mu_{0} r}{2 \pi R^{2}} I_{C}.
\end{eqnarray}
Young and Freedman \cite[p~972]{young2019university} on arriving to equation \ref{28} assert that “When we measure the magnetic field in this region (between the plates of a capacitor), we find that it really is there and that it behaves just as equation predicts. This confirms directly the role of displacement current as a source of magnetic field”. This statement could reinforce the misinterpretation of the different terms of Maxwell's equations as field generators. The magnetic field between the plates of a capacitor is due entirely to the conduction currents in the wires and on the plate surfaces \cite{french2000maxwell,milsom2020untold} as follows if we calculate the $I_{D neta}$ from Equation \ref{23}.
\begin{eqnarray}
\label{29}
I_{D n e t a}=\varepsilon_{0} \frac{d}{d t} \int_{0}^{r} \mathbf{E} \cdot \mathbf{dA}=\varepsilon_{0} r^{2} \pi \frac{d E}{d t}.
\end{eqnarray}
Since the electric field between the plates is uniform, $E=\sigma/ \varepsilon_{0}$, with $\sigma$ being the surface charge density of the capacitor plates, substituting in Eq.~\ref{29} we obtain
\begin{eqnarray}
\label{31}
I_{D}=r^{2} \pi \frac{d \sigma}{d t}
\end{eqnarray}
where we can conclude that the displacement current is a consequence of the temporal variation of the surface charge density between the plates of the capacitor.

\subsection{\label{magnet}A magnet in motion}

\begin{figure}[h!]
\centering
\includegraphics{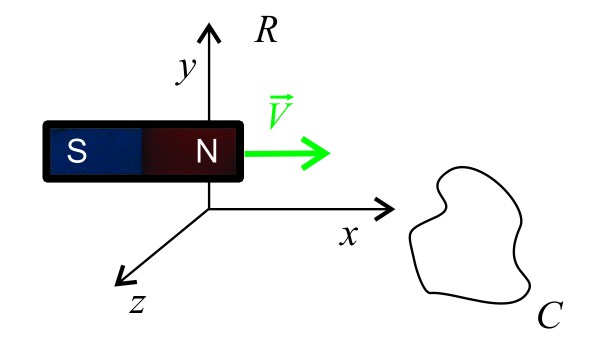}
\caption{Bar magnet performing a URM with respect to an inertial reference frame $R$ and an arbitrary closed curve $C$.}
\label{figure5}
\end{figure}

Another phenomenon widely treated in textbooks refers to electromagnetic induction. Let us consider the case of a bar magnet moving with a constant speed $\boldsymbol{v}=v_{0} \hat{\imath}$ with respect to an inertial frame of reference $R$ and an arbitrary closed curve $C$ as shown in figure \ref{figure5}. 
We know that at all points on the curve there is an electric field induced by the movement of the magnet. This electric field is related to the magnetic field by Faraday's law.
\begin{eqnarray}
\label{32}
\oint \mathbf{E} \cdot \mathbf{dl}=-\frac{d}{d t} \int \mathbf{B} \cdot \mathbf{dA}
\end{eqnarray}
The usual explanation for the appearance of this electric field is that it originates as a consequence of the temporary variation of the magnetic field.

Let us now apply the Ampère-Maxwell law to curve $C$. The magnetic field circulation is related to the temporal variation of the electric field as
\begin{eqnarray}
\label{33}
\oint \mathbf{B} \cdot \mathbf{dl}=\mu_{0} \varepsilon_{0} \frac{d}{d t} \int \mathbf{E} \cdot \mathbf{dA}.
\end{eqnarray}
Following an analogous reasoning to that applied to Faraday's law, we could also say that it is the time-varying electric field that generates the magnetic field. However, both causal relationships between the fields cannot be valid simultaneously. The difficulty is that both fields are generated simultaneously and share the same cause. Both Faraday's and Ampère-Maxwell's laws describe relationships between different fields at the same instant without implying cause-effect relationships.

A magnet moving at a constant speed generates an accompanying electromagnetic field around it. To affirm that the magnetic field generates the electric field is a conceptual simplification, because they are parts of the same entity. 

For both a magnet and a point charge performing a URM with respect to an inertial frame of reference $R$, there is a system $R'$ where they are at rest and only the magnetic or electric aspect of their electromagnetic field is detected. A simple change in the reference system enables us to observe different aspects of the electromagnetic field. We can determine the relation between the fields in $R$ and $R'$ with the Lorentz transformations for \textbf{E} and \textbf{B} or with their weakly relativistic approximation \cite{galili1997changing,ramos2016induccao}. 

\begin{figure}[h!]
\centering
\includegraphics{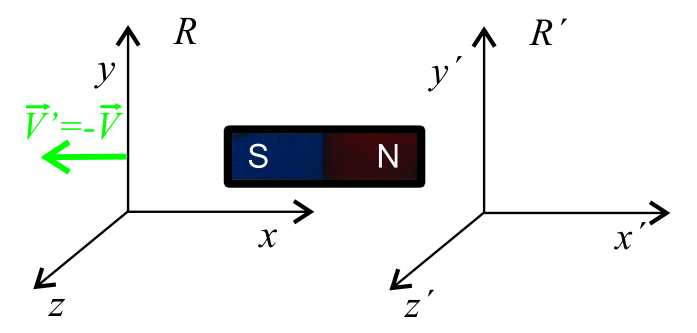}
\caption{Magnet in motion with respect to a reference system $R$ and solidary with respect to the system $R'$.}
\label{figure6}
\end{figure}

Consider for example the case of the magnet. An observer at rest with respect to the magnet in a reference system $R'$ as shown in figure \ref{figure6}, would measure a zero electric field ($\mathbf{E}^{\prime}=\mathbf{0}$), and a non-zero magnetic field $\mathbf{B}^{\prime}=\left(B_{x}^{\prime}, B_{y}^{\prime}, B_{z}^{\prime}\right)$. 
The fields \textbf{E} and \textbf{B} in the systems $R$ and $R'$ are linked by Lorentz transformations \cite{wangsness1986electromagnetic}.
\begin{eqnarray}
\label{35}
\begin{gathered}
E_{x}=E_{x}^{\prime} \\
E_{y}=\gamma\left[E_{y}^{\prime}+v_{0} B_{z}^{\prime}\right] \\
E_{z}=\gamma\left[E_{z}^{\prime}-v_{0} B_{y}^{\prime}\right]
\end{gathered}
\end{eqnarray}
\begin{eqnarray}
\label{36}
\begin{gathered}
B_{x}=B_{x}^{\prime} \\
B_{y}=\gamma\left[B_{y}^{\prime}-(v_{0}/c^2) E_{z}^{\prime}\right] \\
B_{z}=\gamma\left[B_{z}^{\prime}+(v_{0}/c^2) E_{y}^{\prime}\right]
\end{gathered}
\end{eqnarray}
where $\gamma=1/\sqrt{1-(v/c)^{2}}$.

As the electric field in $R'$ is null, the fields result 
\begin{eqnarray}
\label{37}
\mathbf{E}=\left(E_{x}, E_{y}, E_{z}\right)=\left(0, \gamma v_{0} B_{z}^{\prime},-\gamma v_{0} B_{y}^{\prime}\right),
\end{eqnarray}
\begin{eqnarray}
\label{38}
\mathbf{B}=\left(B_{x}, B_{y}, B_{z}\right)=\left(B_{x}^{\prime}, \gamma B_{y}^{\prime}, \gamma B_{z}^{\prime}\right).
\end{eqnarray}
Combining Eq.~\ref{38} with Eq.~\ref{37}, we can find an expression for the electric field in $R$ as a function of the magnetic field
\begin{eqnarray}
\label{39}
\mathbf{E}=v_{0} B_{z} \hat{\jmath}-v_{0} B_{y} \hat{k}.
\end{eqnarray}
As $\mathbf{v}=v_{0} \hat{\imath}$ the last expression is equivalent to
\begin{eqnarray}
\label{40}
\mathbf{E}=-\mathbf{v} \times \mathbf{B}.
\end{eqnarray}

Thus, we note that while an observer in the reference system $R'$ detects only the magnetic field generated by the magnet, another located in $R$ measures electric and magnetic fields linked by equation \ref{40}. An analysis of said equation allows us to recognize the absence of cause-effect between its different terms. A moving magnet generates both an electric field and a magnetic field related by equation \ref{40}.

The electric field and the magnetic field are actually components of the same entity, the electromagnetic field. The inclusion in introductory physics courses of the Lorentz transformations, or their weak relativistic approximation, allows us to recognize that these fields do not have an independent existence and they are mathematically related by these transformations, with no cause-effect relationships between them. Just as it makes no sense to think that a component of one vector is the cause of another, neither does it make sense to think that a component of the electromagnetic field be the cause of the other.

\subsection{\label{faradayfields}Faraday's law and the fields generated by a solenoid}

Now, let us analyze a standard electromagnetic induction problem showed in most of the introductory physics textbooks, which consists of determining the electric field in a region of space where a circular solenoid of length $L$, $n$ turns per unit length and radius $R$, whose current varies with time following a certain function $I(t)$, is located.

\begin{figure}[h!]
\centering
\includegraphics{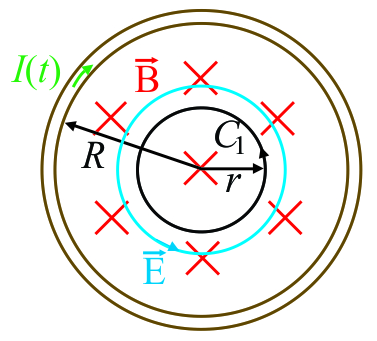}
\caption{Cross section of the solenoid through which a current $I$ is flowing.}
\label{figure7}
\end{figure}

To determine the electric field at a distance $r$ smaller than the radius $R$ of the solenoid, we apply Faraday's law to the closed curve $C_{1}$ illustrated in Fig.~\ref{figure7}, where a cross section is shown. If the solenoid is very long, we can assume that the magnetic field inside it is uniform and the magnetic field flux results as follows
\begin{eqnarray}
\label{Bbobina}
\Phi_B=\mu_{0} n I(t) r^{2} \pi.
\end{eqnarray}
If we substitute Eq.~\ref{Bbobina} in Faraday's law (Eq.~\ref{faraday}) and operate, we obtain the electric field at a distance $r$ from the center of the solenoid
\begin{eqnarray}
\label{Ebobina}
E(t)=\frac{\mu_{0} n r}{2} \frac{d I(t)}{d t}.
\end{eqnarray}
The expression obtained for the electric field is known, emphasizing that it is directly proportional to the time derivative of the current intensity in the solenoid. 
Equations \ref{Bbobina} and \ref{Ebobina} are strictly valid in situations where $I$ is constant or increases linearly with time. Although a discussion about its framework of validity is usually presented in higher electromagnetism courses, if we apply the Ampère-Maxwell's law to the electric field obtained through Faraday's law, we can relatively easily understand its framework of applicability and reveal the way in which Maxwell's equations are linked.

\begin{figure}[h!]
\centering
\includegraphics{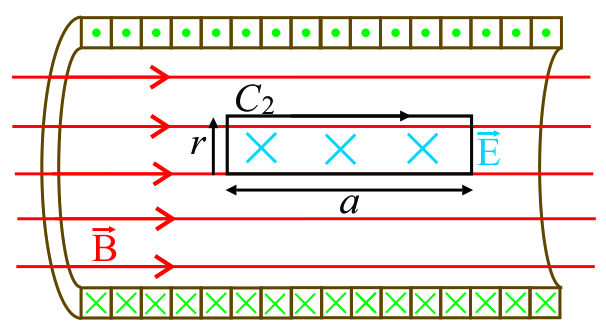}
\caption{Longitudinal section of the solenoid. The magnetic field is horizontal to the right.}
\label{figure8}
\end{figure}

Figure ~\ref{figure8} shows a longitudinal section of the solenoid, together with the directions of the magnetic and electric fields, under the assumption that the current intensity increases with time. If we apply Ampère-Maxwell's law to the closed curve $C_{2}$, since there are no conduction currents through a surface delimiting it, we have that 
\begin{eqnarray}
\label{circulacióncorregida1}
\oint \mathbf{B} \cdot \mathbf{d l}=\mu_{0} \varepsilon_{0} \frac{d}{d t} \int \mathbf{E} \cdot \mathbf{d A}=\mu_{0} \varepsilon_{0} \frac{d \int_{0}^{r} E a d r}{d t}.
\end{eqnarray}
If we substitute Eq.~\ref{Ebobina} in Eq.~\ref{circulacióncorregida1} and operate 
\begin{eqnarray}
\label{circulacióncorregida2}
\oint \mathbf{B} \cdot \mathbf{d l}=\frac{\mu_{0}^{2} \varepsilon_{0} n a r^{2}}{4} \frac{d^{2} I(t)}{d t^{2}}.
\end{eqnarray}
This equation reveals that the magnetic field circulation is directly proportional to the second time derivative of the intensity of the current in the solenoid. In the case in which the magnetic field is uniform, the  circulation along $C_{2}$ should vanish. Then by applying the Ampère-Maxwell law, we directly find the framework of validity of Eqs. \ref{Bbobina} and \ref{Ebobina}. The only way that the electric field obtained through Faraday's law is a solution of Maxwell's equations is that $d^{2} I / d t^{2}=0$, or equivalent, the current intensity varies linearly with time.

The problem posed has several interesting points to discuss, the first of which is the fact that, when dealing with electromagnetic induction problems in introductory physics courses, electric field functions are generally determined using Faraday's law, without checking or discussing whether the solution also verifies Ampère-Maxwell's law. This could lead students to believe that the solutions of the fields only have to verify one of Maxwell's equations. On the other hand, the analysis performed allows us to understand why the equation for the electric field found is only valid when $d^{2} I / d t^{2}=0$. Faraday's and Ampère-Maxwell's laws are coupled, since the electric field circulation depends on the rate of change of the magnetic field flux, and the magnetic field circulation depends on the rate of change of the electric field flux. When we affirm that $d^{2} I / d t^{2}=0$, the rate of change of the electric field flux becomes zero, decoupling Faraday's and Ampère-Maxwell's laws, which then allows us to determine the electric field with Faraday's law alone.

Let us now consider a situation in which $d^{2} I / d t^{2} \neq 0$. In this case, Eq.~\ref{circulacióncorregida2} gives  the first order correction of the magnetic field circulation. A typical approach to the problem of why Faraday's law is not enough to determine the electric field is to argue that if the electric field changes with time it generates a new magnetic field that overlaps with the original one. This would result in another configuration of magnetic fields inside the solenoid that would have to be taken into account to recalculate the electric field and so on. This reasoning, which may be reinforced by an approach of successive approximations to determine the fields, contradicts the way of approaching the problem to find a general solution \cite{batell2003electrically}. To find a solution to the fields, one must solve a system of coupled differential equations for the electromagnetic fields that arises from applying Faraday's and Ampère-Maxwell's laws to the solenoid. Electric and magnetic fields must verify both laws simultaneously at any instant of time. It is not correct to point out that one of the fields is the cause of the other.

\section{\label{discussion}Discussion and implications for teaching}

The discussion raised so far has important implications for the teaching of electromagnetism. A unified approach to electric and magnetic fields, in which the actual sources of the fields are the charge and current densities, either constant or time-varying, would show students a more accurate and probably more consistent picture of field sources, cause-effect relationships, and electromagnetic wave propagation. In this framework, to avoid incorrect interpretations, it is important to adequately present Maxwell's equations. 

Choosing adequate examples it is possible to show that Maxwell's equations do not imply cause-effect relationships, but instead present associations between different magnitudes at the same instant of time.
For example, let us consider the induced currents
in a transformer. By varying the  current intensity in the primary winding
and analyzing by means of the law of
Faraday the currents induced in the secondary, we should state that, if there is an  electric field induced in the secondary, it is because there exists next to it a magnetic field that varies in time and both fields are correlated by this law. Hence, the cause of both fields is the current that varies in time in the primary \cite{rosser2013interpretation, Roche_1987}.

Although the introductory physics textbooks mentioned so far consider that a time-varying magnetic field produces an electric field and vice versa, we find that two of the texts based on PER research, namely \textit{
Matter and interactions} by Chabay and Sherwood \cite{chabay2015matter} and \textit{Six ideas that shaped physics} by Moore \cite{moore2017six}, expound the idea that Maxwell's equations imply associations and not cause-effect relationships. Regarding Faraday's law, Chabay and Sherwood \cite[p~902]{chabay2015matter} state that “The historical term ‘magnetic induction’ is often used to describe this phenomenon, and one says that the time-varying magnetic field ‘induces’ the curly electric field. This is somewhat misleading. It is more correct to say that anywhere we observe a time-varying magnetic field, we also observe a curly electric field. Faraday’s law relates these observations quantitatively". Moore \cite[p~271]{moore2017six}, on the other hand, is more emphatic: “I have been very careful to state that this is the electric field that is correlated with the changing magnetic field, not created by that field. Electromagnetic fields are created only by stationary or moving charged particles\dots~ So though \textbf{E} and \textbf{B} are correlated by Faraday’s law in a given reference frame, correlation is not causation".

Ideally, a unified treatment of the electromagnetic field would avoid conceptual confusions such as those presented in the previous sections. To this end, it is key to develop teaching-learning sequences that allow students to understand that Maxwell's equations do not give information about the sources of the fields but rather describe relationships between different quantities and determine their temporal evolution. This can be done not only in situations where time-varying electric and magnetic fields are present, as in the example given in section.~\ref{faradayfields}, but also when analyzing magnetostatic and electrostatic problems where field expressions are determined by Gauss's or Ampère's laws. Thus, emphasis should be placed on the fact that the fields must simultaneously verify all four Maxwell's equations and not one in particular, reaffirming the fact that they form a system of coupled equations. Along these lines, both in static and time-varying field situations, efforts should be made to discuss problems where hypothetical configurations of electric and magnetic fields are analyzed to verify Maxwell's equations. Another possible approach to Gauss's and Ampère's laws, which can be easily extended to Ampère-Maxwell's law, is to analyze the field line configurations for different distributions of charge and current. These problems would stimulate students to reflect on field sources and avoid the emergence of causal linear reasoning \cite{guisasola2008gauss}. 

Finally, we point out the convenience of designing tutorials that allow students to understand that electric and magnetic fields are not independent but parts of the same object, the electromagnetic field. This unified vision helps us to avoid the appearance of contradictory situations, such as those presented in this article, and to develop a better understanding of electromagnetic phenomena. 

\section*{Acknowledgment}
The authors would like to thank PEDECIBA (MEC, UdelaR, Uruguay) and express their gratitude for the grant Fisica Nolineal (ID 722) Programa Grupos I+D CSIC 2018 (UdelaR, Uruguay). Part of this research was funded by the Spanish government (MINECO\textbackslash FEDER PID2019 -105172RB-I00)

\providecommand{\noopsort}[1]{}\providecommand{\singleletter}[1]{#1}%

\end{document}